\documentclass{aa}

\usepackage{natbib,graphicx}
\bibpunct{(}{)}{;}{a}{}{,} % to follow the A&A style 

\begin{document}

\title{An approximate theory for substructure propagation in clusters}

\author{D. A. Prokhorov\inst{1,2}\and F. Durret\inst{1} }

\offprints{D.A. Prokhorov \email{prokhoro@iap.fr}}

\institute{Institut d'Astrophysique de Paris, CNRS, UMR 7095,
Universit\'{e} Pierre et Marie Curie, 98bis Bd Arago, F-75014
Paris, France
            \and
            Moscow Institute of Physics and Technology,
            Institutskii lane, 141700 Moscow Region, Dolgoprudnii, Russia }

\date{Accepted . Received ; Draft printed: \today}

\authorrunning{D.A. Prokhorov \& F. Durret}

\titlerunning{An approximate theory for subcluster propagation}

\abstract 
% context heading (optional)
{}
% aims heading (mandatory)
{The existence of dark matter can be proved in an
astrophysical context by the discovery of a system in which the
observed baryons and the inferred dark matter are spatially
segregated, such as the bullet cluster (1E0657-558). The full
descriptions of the dark matter halo and X-ray gas substructure
motions are necessary to forecast the location of the dark halo 
from X-ray maps, which can be confirmed by the detection of a
galaxy concentration or by gravitational lensing.}
% methods heading (mandatory)
{We present an analytical hydrodynamic model to determine
the distance between the X-ray and dark-matter components and the
Mach number of the merger shock.}
% results heading (mandatory)
{An approximate solution is given for the problem of the
substructure propagation in merging clusters. A new method to
predict the position of a dark matter halo in clusters, where
there is a separation between the X-ray gas and the dark halo, is
proposed and applied to the clusters 1E0657-558 and Abell 1763.}
% conclusions
{}


\maketitle

\section{Introduction}

The ``unseen" component of the universe (which dominates its
mass), is a long-standing issue in modern cosmology since the need
for dark matter was originally pointed out by Zwicky (1933). In a
typical cluster of galaxies only 2\% of the total mass is in the
form of galaxies, 10-15\% is in the form of hot X-ray emitting gas
and the remainder is dark matter. Modified Newtonian dynamics
(MOND) is an alternative theory that can be used to model
different observations without assuming the existence of dark
matter (Milgrom 1983). However, MOND cannot account for all the
properties of clusters, where the presence of dark matter remains
unavoidable (Gerbal et al. 1992, Pointecouteau \& Silk 2006).

The existence of dark matter can be confirmed by the discovery of
a system in which the observed baryons and the inferred dark
matter are spatially segregated. One type of object where this
separation occurs in is merging galaxy clusters. During the
collision of two clusters, the galaxies are effectively
collisionless particles, while the plasma clouds are highly
collisional and therefore are slowed by ram pressure. If dark
matter particles are also collisionless, as is widely assumed, any
dark matter present in the system would be located near the
galaxies. One such merging cluster system is 1E0657-558 (z=0.296),
the so-called bullet cluster. The cluster 1E0657-558 has two
primary galaxy concentrations separated by 0.72 Mpc on the sky, a
less massive (T=6 keV) western subcluster and a more massive (T=14
keV) eastern main cluster (Markevitch et al. 2002). Both
concentrations have associated X-ray emitting plasma offset from
the galaxies toward the center of the system (Clowe et al. 2006).

Shock waves driven in the intergalactic medium during the merging
of galaxy clusters have been observed in X-ray imaging and
spectroscopy. Chandra observations of 1E0657-558 have revealed a
bow shock propagating in front of a bullet-like gas cloud moving
away from the core of the main cluster. Based on the gas density
jump across the shock front, Markevitch et al. (2002) derived a
Mach number of the shock $\sim$3. The inferred shock velocity of
$\sim$4700~km~s$^{-1}$ has been commonly interpreted as the
velocity of the ``bullet'' subcluster itself (Markevitch et al.
2002; Hayashi \& White 2006; Markevitch 2006, among others).

Numerical simulations have very recently been run by several teams
and succeed quite well in reproducing the main properties of the
bullet cluster, both qualitatively and quantitatively. For
example, Milosavljevic et al. (2007) conclude that the halo
collision velocity need not be the same as the intergalactic gas
shock velocity; their simulation finds that the velocity of the
CDM halo is $\sim 16\%$ lower than that of the shocks. Springel \&
Farrar (2007) derived a shock speed of about 4500~km~s$^{-1}$ but
a subcluster velocity of only 2600~km~s$^{-1}$ in the rest frame
of the system, so the shock wave propagates faster than the dark
matter clump. Randall et al. (2007) combine numerical simulations
with results derived from X-ray, strong and weak lensing, and
optical observations to place an upper limit on the
self-interaction cross-section of dark matter per unit mass
$\sigma /m$, which rules out most of the range previously invoked
to explain inconsistencies between the standard collisionless cold
dark matter model and observations. Bradac et al. (2006) have
reconstructed the mass distribution of this cluster both from weak
and strong lensing data, based on multi band high resolution HST
ACS images. They confirmed that the total mass in this cluster
does not trace the baryonic mass, as already found by Clowe et al.
(2004, 2006).

For the description of the X-ray substructure propagation, we are
going to use the ``piston in a tube" model, described in Sect. 2.
In accordance with this model, an additional retarding force acts
on the X-ray substructure. The analysis of substructure
propagation is considered in Sect. 3 for various cluster
parameters.\\ Several estimates from the X-ray observations
indicate that around 30\% of all rich clusters exhibit
substructure on a scale of 1~Mpc (Forman \& Jones 1994). The
number of rich clusters with substructure is a measure of the
fraction of galaxy clusters that have recently accreted a
significant fraction of their mass (Richstone et al. 1992), which
can be used to determine $\Omega_{M}$.\\ The dynamical time of the
substructure is derived from the approach of Gunn \& Gott (1972)
in Sect. 4. Unless evaluations of the incidence of X-ray
substructures are overestimated or substructure lasts
significantly longer than a gravitational free-fall time in
clusters, the value of the fractional density $\Omega_{M}$ is
larger than 0.5. The retarding force has an effect on the X-ray
substructure motion and the duration of the merger can be longer
than a gravitational free-fall time because of this force.

We also apply the ``piston in a tube" model to the cluster Abell
1763, where the X-ray gas substructure does not seem to have gone
beyond the main cluster center. A new method to predict the
position of the dark matter halo in clusters, where the observed
baryons and the inferred dark matter are spatially segregated, is
proposed.

\section{The ``piston in a tube" model}

The discovery of a galaxy cluster in which the observed X-ray
plasma and the inferred dark matter are spatially segregated
requires the existence of a retarding force acting on the X-ray
substructure. The observed shock front (in 1E0657-558) moving
ahead of the X-ray clump shows supersonic motion of the gas.

We develop a simple 1-D ``piston in a tube" model to derive the
analytical expression for the retarding force. The retarding force
acting on the X-ray clump (piston) in this model arises because
the pressure of the compressed gas ahead of the piston is higher
than the pressure of the rarefied gas behind the piston. Different
regions of the flow are essential for this model: 1) the region of
calm gas ahead of the piston, 2) the region of shocked gas, 3) the
region of rarefied gas, and 4) the region of calm gas behind the
rarefied gas. The piston is situated at the boundary of regions 2,
3 and moves towards region 1. The shock is situated at the
boundary of regions 1, 2.

In regions of calm gas, 1 and 4, the velocity of the gas equals 0,
the pressure is that of the ambient medium. In region 2
(post-shock region), the velocity of the gas equals the piston
velocity U, the pressure is higher than the initial one. In region
3, the rarefaction wave arises; the velocity of the gas depends on
the distance from the piston and the value of the pressure is
smaller than the initial one. Qualitative sketches of the gas
velocity and pressure profiles in different regions are plotted in
Figs. 1 and 2, respectively.

 \begin{figure}[h]
 \centering
 \includegraphics[angle=0,width=8cm]{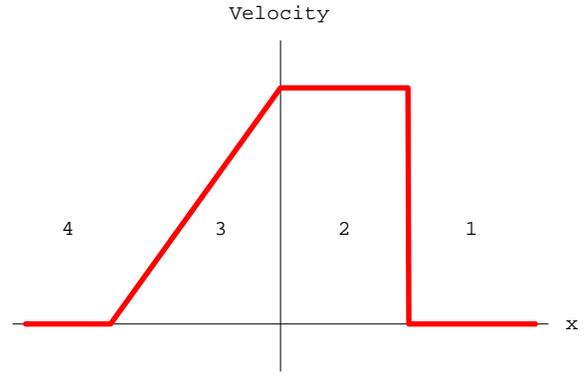}
 \caption{Gas velocity in different regions of the flow. The piston
moves from left to right.}
 \end{figure}

 \begin{figure}[h]
 \centering
 \includegraphics[angle=0,width=8cm]{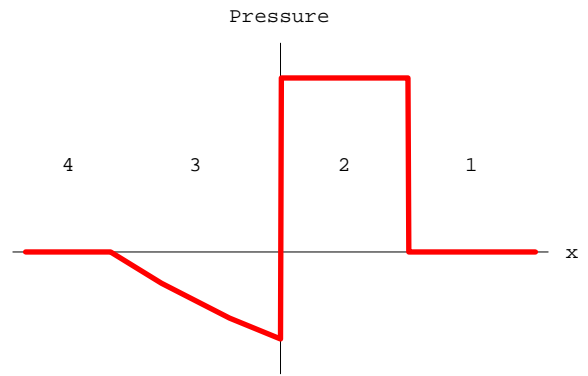}
 \caption{Gas pressure in different regions of the flow.}
 \end{figure}

\subsection{The pressure in front of the substructure\\ (region 2)}
Following Landau \& Lifshitz (1959) we calculate the pressure in
front of the substructure (region 2). The shock wave arises in
front of the substructure (which acts as a piston) because the
substructure has a velocity higher than the sound velocity in the
main cluster. At first the position of the shock coincides with
the position of the piston. Later the shock wave outdistances the
piston and a region of shocked gas appears between the shock and
the substructure. The gas pressure in front of the shock wave
(region 1) equals the initial pressure $p_{1}$ and its velocity
equals zero. The gas in the region between the shock and the
substructure (region 2) has the substructure velocity. The
difference between the velocities of regions 1 and 2 equals the
substructure velocity U.

The conditions on the shock (continuity of matter flux, momentum
flux, energy flux) can be written as:
\begin{eqnarray}
&&\rho_{1} v_{1}=\rho_{2} v_{2}\equiv j\\
&&p_{1}+\rho_{1}v^2_{1}=p_{2} +\rho_{2} v^2_{2}\\
&&\frac{v^2_{1}}{2}+w_{1}=\frac{v^2_{2}}{2}+w_{2}
\end{eqnarray}
where $v$ is the gas velocity in the  reference frame of the
shock,
$\rho$ is the gas density, $p$ is the pressure, and $w$ is the specific enthalpy.\\
From the first equation we find the difference of the velocities:
\begin{equation}
v_{1}-v_{2}=\frac{j}{\rho_{1}}-\frac{j}{\rho_{2}}=j\cdot(V_{1}-V_{2})
\end{equation}
where $V$ is the inverse of the gas density. \\
Therefore we can express the matter flux
\begin{equation}
j=\frac{v_{1}-v_{2}}{V_{1}-V_{2}}
\end{equation}
Using the expression $v_{1,2}=j\cdot V_{1,2}$ and Eq. (2), we can
find
\begin{equation}
p_{1}+j^2 V_{1}=p_{2}+j^2 V_{2}
\end{equation}
One can express $j^2$:
\begin{equation}
j^2=\frac{p_{2}-p_{1}}{V_{1}-V_{2}}
\end{equation}
We can use the Eqs. (5) and (7) together to calculate the velocity
U of the substructure:
\begin{equation}
U\equiv v_{1}-v_{2}=\sqrt{(p_{2}-p_{1})(V_{1}-V_{2})}
\end{equation}
Using the expression $v_{1,2}=j\cdot V_{1,2}$ and Eq. (3), we can
find
\begin{equation}
w_{1}+\frac{j^2V^2_{1}}{2}=w_{2}+\frac{j^2 V^2_{2}}{2}
\end{equation}
Using Eq. (7), we can find
\begin{equation}
w_{1}-w_{2}=\frac{1}{2}(V_{1}+V_{2})(p_{1}-p_{2})
\end{equation}
The specific enthalpy of polytropic gas is
\begin{equation}
w_{1,2}=\frac{\gamma p_{1,2} V_{1,2}}{(\gamma-1)}
\end{equation}
where $\gamma=c_{p}/c_{v}$.\\
 After the simple transformation of Eqs. (10) and (11),
 one can find the equation for the percussive adiabat
\begin{equation}
\frac{V_{2}}{V_{1}}=\frac{(\gamma+1)p_{1}+(\gamma-1)p_{2}}{
(\gamma-1)p_{1}+(\gamma+1)p_{2}}
\end{equation}
Substituting $V_{2}$ from Eq. (12) in Eq. (8) one can simplify the
expression for the substructure velocity
\begin{equation}
U=(p_{2}-p_{1})\sqrt{\frac{2V_{1}}{(\gamma-1)p_{1}+(\gamma+1)p_{2}}}
\end{equation}
For a polytropic gas $p=A\rho^{\gamma}$, therefore the sound
velocity:
\begin{equation}
c^2=\frac{\partial p}{\partial\rho}=\gamma A
\rho^{\gamma-1}=\gamma p V
\end{equation}
One can find the expression for the pressure in the region 2 from
Eqs. (13) and (14):
\begin{equation}
\frac{p_{2}}{p_{1}}=1+\frac{\gamma(\gamma+1)U^2}{4c^2_{1}}+\frac{\gamma
U}{c_{1}}\sqrt{1+\frac{(\gamma+1)^2U^2}{16c^2_{1}}}
\end{equation}
where $c_{1}$ is the sound velocity in the first region.

\subsection{The pressure behind the substructure (region 3)}
The rarefaction wave is formed behind the substructure (region 3).
The self-similarity solution of the flow is described in Landau \&
Lifshitz (1959), this solution depends on the single parameter
$\xi=x/t$, where $x$ is a distance, t is a time parameter. The
equation of entropy conservation is
\begin{equation}
\frac{\partial s}{\partial t} + v \frac{\partial s}{\partial x}=0
\end{equation}
where s is the specific entropy, $v$ is the flow velocity.\\
The derivatives can be expressed as
\begin{eqnarray}
&&\frac{\partial}{\partial t}=-\frac{\xi}{t}\frac{d}{d\xi}\\
&&\frac{\partial}{\partial x}=\frac{1}{t}\frac{d}{d\xi}
\end{eqnarray}
Using Eqs. (16) , (17), (18) one can find
\begin{equation}
(v-\xi)s^{\prime}_{\xi}=0
\end{equation}
where $s^{\prime}_{\xi}\equiv ds/d\xi$.\\
The Euler equation and the continuity of matter flux equation are
\begin{eqnarray}
&&\frac{\partial v}{\partial t} + v \frac{\partial v}{\partial
x}=-\frac{1}{\rho}\frac{\partial p}{\partial x}\\
&&\frac{\partial \rho}{\partial t}+\rho\frac{\partial v}{\partial
x} + v \frac{\partial \rho}{\partial x}=0
\end{eqnarray}
We can rewrite these equations:
\begin{eqnarray}
&&(v-\xi)\rho^{\prime}_{\xi}+\rho v^{\prime}_{\xi}=0\\
&&(v-\xi)v^{\prime}_{\xi}=-\frac{c^2}{\rho}\rho^{\prime}_{\xi}
\end{eqnarray}
Consequently from Eqs. (19) and (22), we have $s^{\prime}_{\xi}=0$
and $s=const$ and the self-similarity solution is isoentropic.
Eliminating $\rho^{\prime}_{\xi}$, $v^{\prime}_{\xi}$ from these
two equations, one can find the condition for the non-trivial
solution (Landau \& Lifshitz 1959):
\begin{equation}
\xi-v=c
\end{equation}
Using the Eqs. (22) and (24), we have
\begin{equation}
c\rho^{\prime}_{\xi}=\rho v^{\prime}_{\xi}
\end{equation}
which can be integrated to give $v$:
\begin{equation}
v=\int\frac{c(\rho) d\rho}{\rho}
\end{equation}
For adiabatic processes
\begin{equation}
\rho T^{1/(1-\gamma)}=const
\end{equation}
The sound velocity is proportional to the square root of the
temperature, therefore
\begin{equation}
\rho=\rho_{0}\left(\frac{c}{c_{1}}\right)^{2/(\gamma-1)}
\end{equation}
Using Eqs. (26) and (28) we have
\begin{equation}
v=\frac{2}{\gamma-1}\int dc=\frac{2}{\gamma-1}(c-c_{1})
\end{equation}
where $c_{1}$ is the sound velocity in the region of the calm gas.
\begin{equation}
c=c_{1}-\frac{\gamma-1}{2}|v|
\end{equation}
Using Eq. (28), we obtain the density in the region 3:
\begin{equation}
\rho_{3}=\rho_{1}\left(1-\frac{\gamma-1}{2}\frac{|v|}{c_{1}}\right)^{2/(\gamma-1)}
\end{equation}
and the pressure $p_{3}=A\rho^{\gamma}_{3}$:
\begin{equation}
p_{3}=p_{1}\left(1-\frac{\gamma-1}{2}\frac{|v|}{c_{1}}\right)^{2\gamma/(\gamma-1)}
\end{equation}

\subsection{The retarding force}
From Eqs. (15) and (32) the retarding force $F=(p_{3}-p_{2})S$ is:
\begin{eqnarray}
&&F=p_{1}S\left(-\left(1+\frac{\gamma(\gamma+1)U^2}{4c^2_{1}}+\frac{\gamma
U}{c_{1}}\sqrt{1+\frac{(\gamma+1)^2U^2}{16c^2_{1}}}\right)\right.+
\nonumber\\
&&+\left.\left(1-\frac{\gamma-1}{2}\frac{U}{c_{1}}\right)^{\frac{2\gamma}{\gamma-1}}\right)
\end{eqnarray}
Let $\gamma=5/3$ and $u=U/3c_{1}$, we can rewrite Eq. (33) using
these definitions
\begin{equation}
F(u)=p_{1}S\left((1-u)^5-1-10u-5u\sqrt{1+4u^2}\right)
\end{equation}
If $u\ll1$, one can calculate the Taylor series
\begin{equation}
G(u)=-20p_{1}Su
\end{equation}
The functions $F/(p_{1}S)$ and $G/(p_{1}S)$ are shown in Fig. 3.\\

\begin{figure}[h]
\centering
\includegraphics[angle=0, width=8cm]{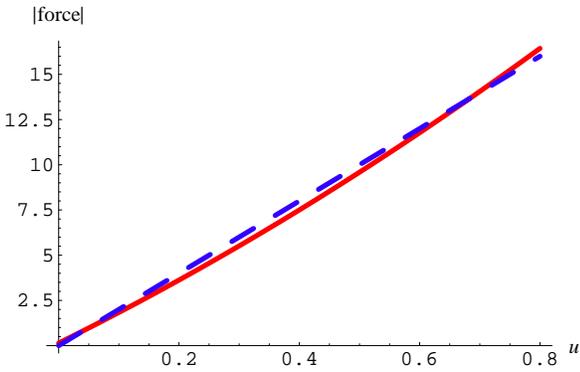}
\caption{Functions $F/(p_{1}S)$ vs $u$ (solid line) and
$G/(p_{1}S)$ vs $u$ (dashed line).}
\end{figure}

Because of the coincidence of these functions, the function $G(u)$
will be useful later. Thus, we can consider that the retarding
function is proportional to the substructure velocity.

\section{Simulation of substructure propagation}

\subsection{Merger initial velocity}
During the merger of two clusters of galaxies, the baryonic
component, feeling the gravitational potential created by the dark
matter components, moves supersonically (Gabici \& Blasi 2003).
Let us assume that two clusters are completely virialized
structures with masses $M_{1}$ and $M_{2}$ ($M_{2}\ll M_{1}$,
e.g., the cluster 1E0657-558). To approximate the merger velocity
of the system, we use the assumption (Berrington \& Dermer 2003)
that the cluster of mass $M_{2}$ is a point mass. The main cluster
accretes another cluster that falls from a turnaround radius
$R_{ta}$ of the system, where the two subclusters are supposed to
have zero relative velocity. The exact value of the turnaround
radius does not strongly affect the collision velocity as long as
it is large and the infall velocity approaches free-fall from
infinity. Using the energy conservation law, one can find the
velocity of the subcluster with mass $M_{2}$ at the virial radius
$R_{vir1}$ of the massive cluster, which has the mass $M_{1}$:
\begin{equation}
\frac{v^2_{0}}{2}=\frac{GM_{1}}{R_{vir1}}-\frac{GM_{1}}{R_{ta}}
\end{equation}

\begin{equation}
v_{0}=\sqrt{\frac{2GM_{1}}{R_{vir1}}-\frac{2GM_{1}}{R_{ta}}}
\end{equation}
The definition of the virial radius is
\begin{equation}
R_{vir1}=\left(\frac{3M_{1}}{4\pi
\Delta_{c}\Omega_{M}\rho_{cr}}\right)^{1/3}
\end{equation}
where $\Delta_{c}$ is the density contrast for the formation of
the cluster, $\Omega_{M}$ is the matter density fraction and the
critical density is
\begin{equation}
\rho_{cr}=\frac{3 H^2}{8\pi G}
\end{equation}

\subsection{Solution without taking the tidal force into account}
The gas number density profiles of X-ray clusters of galaxies can
be approximated by the empirical isothermal $\beta$ model
(Cavaliere \& Fusco-Femiano 1976)
\begin{equation}
n(r)=n_{0}\left(1+\left(\frac{r}{r_{c}}\right)^2\right)^{-3\beta/2}
\end{equation}
where $n_{0}$ is the central gas number density, $r_{c}$ is the
core radius of the cluster and $\beta$ is the beta parameter.

Recent high-resolution N-body simulations have suggested that dark
halos of clusters are described by a family of universal density
profiles. Navarro et al. (1997) proposed the following profile
(NFW profile):
\begin{equation}
\rho_{DM}(r)=\frac{\delta_{c}\rho_{cr}\Omega_{M}}{(r/R_{s})(1+r/R_{s})^2}
\end{equation}
where the characteristic dimensionless density is
\begin{eqnarray}
&&\delta_{c}=\frac{200}{3}\frac{c^3}{\ln(1+c)-c/(1+c)}\\
&&R_{s}=R_{vir}/c\nonumber
\end{eqnarray}
where c is a concentration parameter (Navarro et al. 1997).

It is useful to introduce a dimensionless time
\begin{equation}
\tau=t\frac{c_{s}}{R_{vir1}}
\end{equation}
The initial conditions $x(0), \tilde{v}(0)$ are
\begin{eqnarray}
&&x(0)=-R_{vir1}\\
&&\tilde{v}(0)=\frac{dx}{d\tau}(0)=v_{0}\frac{R_{vir1}}{c_{s}}\nonumber
\end{eqnarray}
where $\tilde{v}\equiv dx/d\tau$.

The gravitational force acting on the X-ray gas of the subcluster
equals
\begin{eqnarray}
&&F_{gr}(R)=\frac{G M_{gas 2}}{R^2}\int^{R}_{0}4\pi x^2
\rho_{DM}(x) dx\\
&&F_{gr}(R)=\frac{4\pi G M_{gas 2}\delta_{c}\rho_{cr} \Omega_{M}
R^3_{s}}{R^2}\times\\
&&\left(\ln\left(1+\frac{R}{R_{s}}\right)-\frac{R/R_{s}}{1+R/R_{s}}\right)\nonumber
\end{eqnarray}
where $M_{gas 2}$ is the gas mass of the X-ray subcluster.

The equation of motion of the X-ray gas of the substructure
without taking into account the tidal force (only the
gravitational Eq. 46 and retarding forces Eq. 35 are taken into
account) is then:
\begin{eqnarray}
&&\frac{d^2 x}{d\tau^2}+\frac{20}{3}\frac{n(x)kTSR_{vir1}}{M_{gas
2}c^2_{s}}\frac{dx}{d\tau}+ 4\pi G\delta_{c}\rho_{cr} \Omega_{M}
R^3_{s}\times\\&&
\times\frac{R^2_{vir1}}{c^2_{s}}\left(\ln\left(1+\frac{|x|}{R_{s}}\right)-
\frac{|x|}{R_{s}}\left(1+\frac{|x|}{R_{s}}\right)^{-1}\right)\cdot
\frac{|x|}{x^3}=0\nonumber
\end{eqnarray}

The equation of motion of the dark (collisionless) matter
substructure is (no retarding force):
\begin{eqnarray}
&&\frac{d^2 y}{d\tau^2}=-4\pi G \delta_{c}\rho_{cr} \Omega_{M}
R^3_{s} \frac{R^2_{vir1}|y|}{c^2_{s}y^3}\times\\
&&\left(\ln\left(1+\frac{|y|}{R_{s}}\right)-
\frac{|y|}{R_{s}}\left(1+\frac{|y|}{R_{s}}\right)^{-1}\right)\nonumber
\end{eqnarray}

\subsection{The characteristic parameter of the problem}
The second term on the left-hand side of Eq. (47) weakly depends
on the space coordinate in the cluster core. When $x\ll R_{s}$,
the third term on the left-hand side of Eq. (47) can be
simplified. One can derive the Taylor series
\begin{equation}
\ln\left(1+\frac{|x|}{R_{s}}\right)-\frac{|x|}{R_{s}}\cdot\left(1+\frac{|x|}{R_{s}}\right)^{-1}\sim
\frac{x^2}{2R^2_{s}}+o\left(\frac{x^2}{R^2_{s}}\right)
\end{equation}

The simplified expression of Eq. (47) is
\begin{equation}
x^{\prime\prime}+\frac{20n_{0}kTSR_{vir1}}{3 M_{gas
2}c^2_{s}}x^{\prime}+ \frac{4\pi G
R^2_{vir1}\delta_{c}\rho_{cr}\Omega_{M}}{2 c^2_{s}}x=0
\end{equation}
This equation is the same as the harmonic oscillator one
\begin{equation}
x^{\prime\prime}+2\lambda x^{\prime}+\omega^2_{0}x=0
\end{equation}
The characteristic parameter of such an oscillator motion is
\begin{equation}
\frac{\lambda}{\omega_{0}}=\frac{10n_{0}kTS}{3M_{gas 2}c_{s}}
\sqrt{\frac{1}{2\pi G\delta_{c}\rho_{cr}\Omega_{M}}}
\end{equation}
The dependence of the X-ray substructure motion with time, which
is obtained from Eq. (47) is shown in Fig. 4 for the different
values of the characteristic parameter ($\lambda/\omega_{0}=0.25,
0.5, 1$).\\ The value of $\lambda/\omega_{0}=0.5$ corresponds to
the first set of parameters, i.e., to 1E0657-56 (see Table 1).

 \begin{figure}[h]
 \centering
 \includegraphics[angle=0, width=8cm]{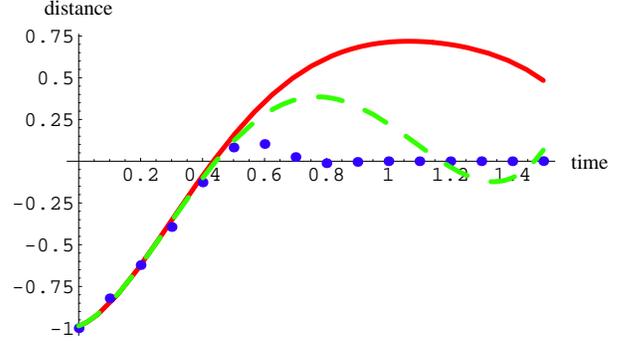}
 \caption{Dependence of the X-ray substructure position $x/R_{v}$ on the
dimensionless time $t R_{v}/c_{s}$ for different characteristic
parameters $\lambda/\omega_{0}$=0.25 (solid line), 0.5 (dashed
line), 1.0 (dotted line).}
\end{figure}

The characteristic parameter is useful for a classification of the
X-ray substructure motion by analogy with an oscillator. If
$\lambda/\omega_{0}<1$ the oscillator motion is a dying
oscillation, while if $\lambda/\omega_{0}>1$ the oscillator motion
is an aperiodic damping.

\subsection{Solution taking the tidal force into account}
It is necessary to consider the influence of the substructure dark
matter halo motion on the X-ray substructure motion. The
assumption that the small cluster (the dark matter halo and the
gas substructure) is a point mass is not acceptable because of the
infinite gravitational energy at the initial time. It is a
reasonable approximation to calculate the tidal force by assuming
a NFW profile for the dark matter substructure halo. The tidal
force acting on the subcluster dark halo is much smaller than the
gravitational force between the two dark halos, because the tidal
force is proportional to the mass of the X-ray gas of the
subcluster. Consequently the dark matter substructure motion
$y(\tau)$ is described by Eq. (48). However, we must include the
tidal force in Eq. (47) for the motion of the X-ray substructure
by analogy with Eq. (46).
\begin{eqnarray}
&&F_{tidal}=4\pi M_{gas 2}\frac{R^2_{vir1}}{c^2_{s}}
\delta_{c}\rho_{cr} \Omega_{M}
R^3_{s2}\frac{|z-y|}{(y-z)^3}\times\\
&&\left(\ln\left(1+\frac{|z-y|}{R_{s2}}\right)-\frac{|z-y|}{R_{s2}}
\left(1+\frac{|z-y|}{R_{s2}}\right)^{-1}\right)\nonumber
\end{eqnarray}
where z is the position of the X-ray subcluster relative to the
main cluster center in the case where the tidal force is taken
into account.

\subsection{Mach number}
One task is to find the Mach number in the pre-shock region. This
is used to determine the temperature ratio of the post-shock
region to that in the pre-shock region. One can obtain the
dependence of the Mach number in the pre-shock region on the Mach
number of the piston:
\begin{equation}
\frac{v_{2}}{v_{1}}=\frac{v_{sh}-U}{v_{sh}}=\frac{(\gamma-1)M^2_{1}+2}{(\gamma+1)M^2_{1}}
\end{equation}
where $v_{2}$, $v_{1}$ are velocities of the post-shock and
pre-shock regions. Consequently, in the frame reference of the
shock, $v_{sh}=c_{1}M_{1}$ is the velocity of the shock, U is the
velocity of the piston.\\
With $\gamma=5/3$, one can find from Eq. (54):
\begin{equation}
\frac{U}{c_{1}}=\frac{3}{4}\frac{M^2_{1}-1}{M_{1}}
\end{equation}
Therefore the Mach number of the piston is smaller than the Mach
number in the pre-shock region.

\subsection{Examples: 1E0657-56 and Abell 1763}

The two sets of selected cluster parameters are in Table 1:\\
1) For the first type of mergers when the substructure has passed
through the main cluster core, the data on the galaxy cluster
1E0657-56 (the ``bullet" cluster) are used (Markevitch, private
communication)\\
2) For the second type of mergers when the substructure has not
gone beyond the main cluster center, as in the cluster
of galaxies Abell 1763 (Lima Neto \& Durret. 2007).\\

Table 1. Sets of parameters for 1E0657-56 and Abell 1763\\

\begin{tabular}{|l|l|l|}
\hline  & 1E0657-56 & Abell~1763 \\
\hline Cluster mass $M_{1}$, $10^{15}M_{\bigodot}$ & 2 & 0.5\\
\hline Central density $n_{0}$, ($cm^{-3}$) & $0.005$ & $0.0079$\\
\hline Core radius $r_{c}$, (kpc) & 180 & 136\\
\hline Mass ratio $M_{1}/M_{2}$ & 60 & 22\\
\hline Subcluster density, ($cm^{-3}$) & 0.02& 0.0079\\
\hline Time unit $R_{vir1}/c_{s}, 10^{9}yr$ &3.4&2.8\\
\hline $\lambda/\omega_{0}$&0.5&3.1\\ \hline
\end{tabular}\\

where the NFW concentration parameter is 4 in both cases.

\subsection{1E0657-56}

\begin{figure}[h]
\centering
\includegraphics[angle=0, width=8cm]{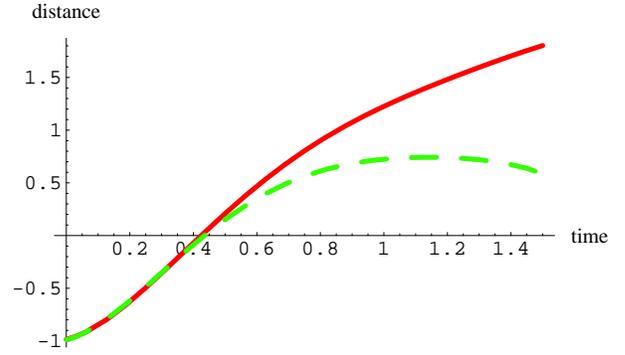}
\caption{Dependence of the X-ray substructure position
$z/R_{vir1}$ (dashed line) and of the dark halo position
$y/R_{vir1}$ (solid line) on the dimensionless time $t
R_{vir1}/c_{s}$ for the first set of parameters (1E0657-56).}
\end{figure}

The results can be compared with those of Clowe et al. (2006). The
cluster exhibits a prominent bow shock with a Mach number
$M=3.0\pm0.4$. The distance observed between the two centers,
corresponding to the X-ray and dark-matter components of the
smaller subcluster, is around 200 kpc, and the distance between
the centers of the dark-matter substructures is around 1 Mpc.

We approximate the shape of the X-ray subcluster with a cylinder
(radius R=79 kpc and length L=120 kpc). For the first set of
parameters, the position of the X-ray substructure is derived from
generalized Eq. (47) taking the tidal force given in Eq. (53) into
account, and the position of the dark halo is calculated from Eq.
(48); these quantities are plotted in Fig.~5.

The dimensionless time $t R_{v}/c_{s}=0.51$ corresponds to the
observed distance of 1~Mpc between the centers of the two
dark-matter substructures and is used to determine the Mach number
of the shock and the distance between the X-ray and dark-matter
components.

The Mach number of the X-ray substructure $dz/(c_{s} dt)$ is shown
in Fig. 6 and the distance between the X-ray and dark-matter
components $y(\tau)-z(\tau)$ is shown in Fig. 7, as functions of
the dimensionless time $t R_{v}/c_{s}$.

From the condition $t R_{v}/c_{s}=0.51$, we find the value of the
Mach number of the X-ray substructure to be $U/c_{1}=2$ from
Fig.~6 and the distance between the X-ray and dark-matter
components to be 200~kpc from Fig.~7. The Mach number of the shock
$M_{1}=3$ is found by solving the quadratic Eq. (55).

Numerical hydrodynamical simulations of the bullet cluster
1E0657-558 (Springel \& Farrar 2007; Milosavljevic et al. 2007)
reached a very similar conclusion to ours: the halo collision
velocity need not be the same as the intergalactic gas shock
velocity, and while the kinematics of the shock are sensitive to
the details of the cluster structure, the instantaneous shock
velocity can exceed the relative velocity of CDM halos. 2D
numerical simulations provide much more precise results than our
approximate theory, because we made several simplifying
assumptions (e.g., the problem is 1-D), and for accurate
description of merging clusters numerical simulations are very
useful. On the other hand, numerical simulations provide
information for very specific sets of parameters. One immediate
question is what are the key controlling parameters and might they
affect the results? For example, using different simulation
parameters Milosavljevic et al. (2007) and Springel \& Farrar
(2007) find that the velocities of the CDM halos are respectively
$\sim$16\% and $\sim$40\% lower than that of the shock.
Milosavljevic et al. (2007) strongly emphasize that the shock
kinematics and morphology are extraordinarily sensitive to the
parameters of the simulations. The advantage of our approximative
theory is that it has a single characteristic parameter
$\lambda/\omega_{0}$, which completely determines the X-ray
substructure motion.

The similar results of Milosavljevic et al. (2007) and Springel \&
Farrar (2007), i.e., that the shock velocity is higher than the
dark halo velocity, are expected for merging clusters, where the
characteristic parameter $\lambda/\omega_{0}$ is much smaller than
1. The influence of the retarding force in these systems is
negligible and the velocity of the shock is higher than those of
the X-ray substructure and the dark halo, because the shock
outdistances the piston in the ``piston in a tube" model.

 \begin{figure}[h]
 \centering
 \includegraphics[angle=0, width=8cm]{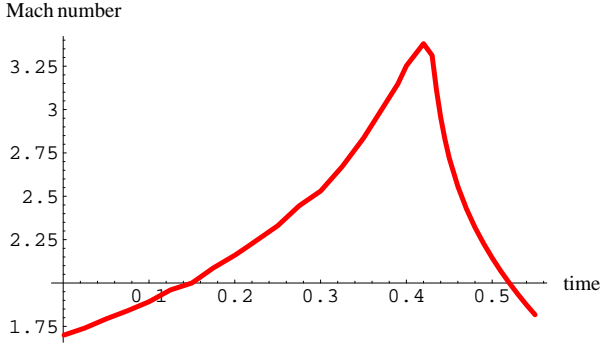}
 \caption{Mach number of the X-ray structure vs the dimensionless time $t R_{vir1}/c_{s}$
 for 1E0657-56.}
 \end{figure}

 \begin{figure}[h]
 \centering
 \includegraphics[angle=0, width=8cm]{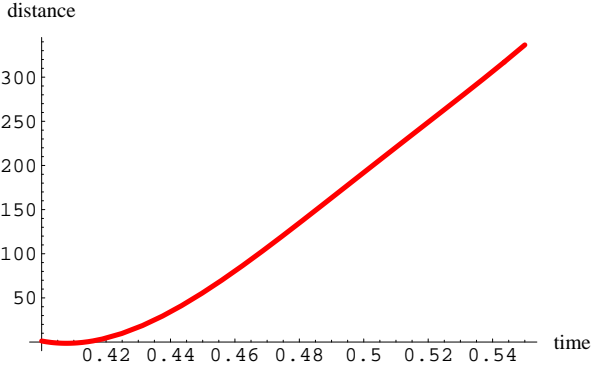}
 \caption{Dependence of the distance (kpc) between the X-ray and dark-matter
 substructures on the dimensionless time $t R_{vir1}/c_{s}$ for 1E0657-56.}
 \end{figure}

\subsection{Abell 1763}
We assume that the X-ray subcluster has the shape of a sphere and
the radius of the X-ray subcluster equals the radius of the
hottest region in the cluster, that is $R\approx140$ kpc.
\begin{figure}[h]
\centering
\includegraphics[angle=0, width=8cm]{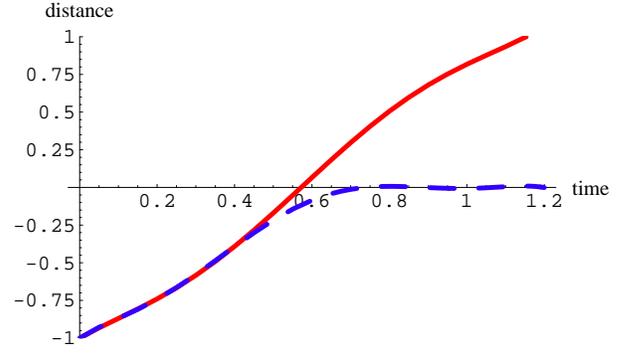}
\caption{Dependence of the X-ray substructure position
$z/R_{vir1}$ (dashed line) and of the dark halo position
$y/R_{vir1}$ (solid line) on the dimensionless time $t
R_{vir1}/c_{s}$ for the second set of parameters (Abell 1763).}
\end{figure}
We propose that the hottest region in Abell~1763 (where the
temperature is twice the average temperature) is situated in the
post-shock region. From the Rankine-Hugoniot condition we can
estimate the Mach number in the pre-shock region. The ratio of the
temperatures in the two regions is:
\begin{equation}
\frac{T_{2}}{T_{1}}=\frac{\left(2\gamma
M^2_{1}-(\gamma-1)\right)\left((\gamma-1)M^2_{1}+2\right)}{(\gamma+1)^2
M^2_{1}}
\end{equation}
Using Eq. (55) we find the Mach number of the piston to be
$U/c_{1}=1.1$. One can see from Fig. 9 that the dark matter
substructure has a much higher velocity than the X-ray
substructure. Therefore the effect of the retarding force is
stronger in this cluster than in 1E0657-56. One can therefore
expect a separation of the dark matter halo of the substructure
and the X-ray substructure (Fig. 10). If the galaxies and X-ray
subclump are not segregated,
 a galaxy concentration near the region of the substructure
dark matter halo is expected about 200 kpc (time = 0.6) from the
center of the cluster Abell~1763 in the direction of the merger
(south east to north west).

\begin{figure}[h]
\centering
\includegraphics[angle=0, width=8cm]{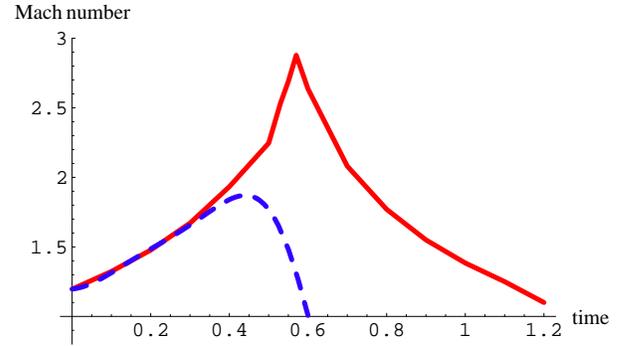}
\caption{Mach number of the piston (dotted line) and of the dark
halo (solid line) vs dimensionless time $t R_{vir1}/c_{s}$ for
Abell 1763.}
\end{figure}
\begin{figure}[h]
\centering
\includegraphics[angle=0, width=8cm]{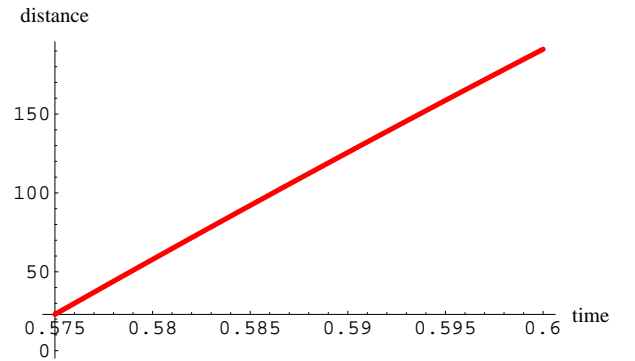}
\caption{Position of the dark matter substructure halo (kpc) vs
dimensionless time $t R_{vir1}/c_{s}$ for Abell 1763.}
\end{figure}

In order to check this point, we analyzed an optical image of
Abell~1763 in the R band taken with the LFC camera on the Palomar
telescope and kindly provided by D.~Fadda and A.~Biviano. The
SExtractor software was applied to the image, and stars were
discarded on the basis of a plot of effective radius as a function
of magnitude, leading to a catalogue of galaxies. From a magnitude
histogram, the galaxy catalogue was estimated to be complete up to
R $\sim$24. We therefore truncated the catalogue at this magnitude
to avoid effects due to incompleteness.

Galaxies from this catalogue were then counted in squares of
140$\times$140~kpc$^2$ centered on the cluster center. The numbers
of galaxies in each square were found to be between 24 and 52,
with the lowest number in the NW region, where our model predicts
an overabundance of galaxies. The contamination by background
galaxies is estimated to be 9 galaxies in each square, as derived
from the AAO model by Bland-Hawthorn \& Ellis
(http://www.aao.gov.au/astro/GalaxyCount/).

There is therefore no galaxy excess 200~kpc NW of the cluster
center, indicating that the X-ray gas of the subcluster must have
stopped while the galaxies have had time to travel further out.
Hence, a direct correspondence between the hottest region in
Abell~1763 and the post-shock region is not correct. Abell~1763 is
therefore a structure in which the observed baryons of the
subcluster and the galaxy concentration are spatially segregated;
a full analysis of the optical image will be done elsewhere and is
beyond the scope of this paper.

Dark matter annihilation in the halos of galaxy clusters has
relevant astrophysical implications (e.g., Colafrancesco 2006). In
fact, if dark matter is constituted by weakly interacting massive
particles (WIMPs), their annihilation can produce direct
observable signals. However, the spatial and spectral intensities
of the astrophysical signals coming from $\chi\chi$ annihilation
are expected to be confused with, or even overcome by, other
astrophysical signals originating from the intracluster gas and/or
from the relativistic plasmas present in the cluster atmospheres,
especially when all these components are co-spatially distributed
with the DM component. An ideal system to detect a DM annihilation
signal would therefore be a cluster with a clear spatial
separation between the various matter components (Colafrancesco et
al. 2007).

\section{Evaluation of the incidence of substructures}

Clusters of galaxies form by the mergers of smaller clusters and
groups. The presence of substructure in clusters of galaxies can
be used to estimate the cosmological density parameter
$\Omega_{M}$.

X-ray data allow the quantification of substructures because
cluster mergers compress and heat the intracluster gas, and this
can be measured as distortions of the spatial distribution of
X-ray surface brightness and temperature (Forman \& Jones 1994;
Jones \& Forman 1999). An alternative method at optical
wavelengths to study substructures is based on the distribution of
cluster galaxies. However, the optical method requires a large
number of galaxies (Dutta 1995), at least a few hundred, and is
more susceptible to contamination from foreground and background
objects.

X-ray studies of cluster substructure use a number of different
statistics (see Buote 2002 for a review). Several studies have
used the ellipticity (Gomez et al. 1997; Kolokotronis et al. 2001;
Melott et al. 2001; Plionis 2002), while other studies measured
centroid (or center-of-mass) variations (Mohr et al. 1995; Gomez
et al. 1997; Kolokotronis et al. 2001). Another method developed
by Buote \& Tsai (1995) is the power ratio method, which is
constructed from the moments of the X-ray surface brightness.

Richstone et al. (1992) performed the first theoretical study of
the relationship of substructure to cosmology. They assumed that
any substructure is wiped out in a cluster crossing time and
calculated the fraction of clusters in the spherical growth
approximation that formed within the last crossing time as a
function of $\Omega_{M}$ and $\Lambda$. This fraction primarily
depends on $\Omega_{M}$, and they estimated that $\Omega_{M}>0.5$
based on estimates of the frequency of substructure in
low-redshift clusters from X-ray images.

The approach proposed by Richstone et al. (1992) to determine
$\Omega_{M}$ suffers several limitations. In their semi-analytical
calculations they avoided the issue of the power spectrum by
concentrating on clusters having the same total mass. The collapse
time of a $10^{15} M_{\bigodot}$ spherical density perturbation
was defined to be the dividing point between clusters that do and
do not possess substructure. The relationship between the collapse
time of a spherical density perturbation and subclustering, though
qualitatively reasonable, is ambiguous. Consequently, they predict
an ambiguous ``frequency of substructure'' rather than a
well-defined quantitative measure of cluster morphology.

A more recent semi-analytical approach was developed by Buote
(1998). He assumed that the amount of substructure depends on the
amount of mass accreted by a cluster over a relaxation timescale
and relates the fractional accreted mass to the power ratios.
Although these semi-analytical methods give an indication of the
expected evolution of cluster substructure and its dependence on
cosmological parameters, perhaps the best method to constrain
cosmological models is through the comparison to cluster
simulations.

Evrard et al. (1993) employed N-body/3D gas dynamic simulations of
the formation of galaxy clusters and demonstrated the existence of
a morphology-cosmology connection for X-ray clusters. Confirming
the analytical expectations of Richstone et al. (1992), they
showed that cluster evolution is sensitive to the cosmological
model in which the clusters form. Numerical simulations show that
both the centroid shift and the power ratios are capable of
distinguishing cosmological models (Evrard et al. 1993; Buote \&
Xu 1997; Valdarnini et al. 1999; Suwa et al. 2003). Buote \& Xu
(1997) find that the power ratios of their ROSAT clusters indicate
an $\Omega_{M}<1$ universe. Valdarnini et al. (1999) find that the
$\Lambda$CDM model is inconsistent with the data. However, they
used $\sigma_{8}=1.1$, which is fairly high and may cause the
disagreement between the $\Lambda$CDM model and the data. Suwa et
al. (2003) find that the center shifts and power ratios of both
surface brightness and projected mass density are able to
discriminate between the cosmological models at z=0.

We uses the value of $\Omega_{M}=0.268\pm0.018$, which was
obtained from the WMAP three year observations (Spergel et al.
2007), to find the dynamical time of the substructure, which is
higher than the gravitational free-fall time. A plausible reason
for this disagreement is that the fraction of clusters with X-ray
substructures exceed the number of clusters that have collapsed
during the last gravitational free-fall time due to the retarding
force acting on the X-ray substructure.

\subsection{The collapse time for bound perturbations}
The separation of two observers in a Friedman-Robertson-Walker
universe with a cosmology constant obeys the equation of motion
\begin{equation}
\frac{d^2 r}{dt^2}=-\frac{4\pi
G\rho_{\star}r^3_{\star}}{3r^2}+\frac{\Lambda r}{3}
\end{equation}
where $\rho_{\star}, r_{\star}$  are the mean density of the
universe and r is the separation between two observers at any specified epoch in the universe. \\
This equation also describes the evolution of the central distance
of a point anywhere in a homogeneous perturbation.

The equation of mass conservation shows that $\Omega_{M}(z)$ and
$\lambda(z)$ of the universe evolve as
\begin{eqnarray}
&&\Omega_{M}(z)=\frac{(1+z)^3\Omega_{M0}}{H^2(z)/H^2_{0}}\\
&&\lambda(z)=\frac{\lambda_{0}}{H^2(z)/H^2_{0}}\nonumber
\end{eqnarray}
where $\Omega\equiv8\pi G\rho/(3H^2)$, $\lambda=\Lambda/(3H^2)$.

The dependence of the rate of expansion H(z) is given by
\begin{equation}
\frac{H^2(z)}{H^2_{0}}=(1-\Omega_{M0}-\lambda_{0})(1+z)^2+\Omega_{M0}(1+z)^3+\lambda_{0}
\end{equation}
It is possible to rewrite Eq. (57) in integral form and estimate
the time to expand from $r_{\star}$ to r for a homogeneous
perturbation:
\begin{equation}
T=H^{-1}_{\star}\int^{r/r_{\star}}_{1}\left(1-\Omega_{M\star}-
\lambda_{\star}+\Omega_{\star}u^{-1}+\lambda_{\star}u^2\right)^{-1/2}du
\end{equation}
Eq. (60) is only valid for perturbations that expand at least as
far as $r/r_{\star}$. Bound perturbations which do not expand
forever reach a maximum expansion and recollapse on a time scale
given by
\begin{equation}
\tau=2T(r_{m}/r_{\star})
\end{equation}
where $r_{m}$ is the first real root $>1$ of the equation:
\begin{equation}
\left(1-\Omega_{M\star}-\lambda_{\star}\right)+\Omega_{M\star}
u^{-1}+\lambda_{\star}u^2=0
\end{equation}

\subsection{Initial density perturbations}
Now suppose that at the epoch defined by $1+z_{1}=1000$ the
universe is characterized by an unperturbed Hubble flow and
cosmological constant, but that the density, on the mass scales of
clusters of galaxies, is normally distributed about the mean
density with variance $\sigma^2$. Thus, we can write for the
perturbations $\delta$:
\begin{eqnarray}
&&\lambda_{\star}=\lambda(z_{1})\\
&&\Omega_{M\star}=\Omega_{M}(z_{1})\cdot(1+\delta)\nonumber
\end{eqnarray}
and $\delta$ has the Gaussian distribution
\begin{equation}
dF(\delta)=\frac{1}{\sqrt{2\pi\sigma^2}}\exp\left(-\frac{\delta^2}{2\sigma^2}\right)d\delta
\end{equation}
The bound perturbation of the density must exceed the critical
value $\delta_{cr}$, which is given by
\begin{eqnarray}
&&1-\Omega_{M}(z_{1})\cdot(1+\delta_{cr})-\lambda(z_{1})+\\
&&+1.5\cdot[2\lambda(z_{1})
\cdot\left(\Omega_{M}(z_{1})\cdot(1+\delta_{cr})\right)^2]^{1/3}=0\nonumber
\end{eqnarray}
One can calculate from Eq. (65) the critical value of perturbations  $\delta_{cr}=2.5\cdot10^{-3}$. \\
The maximum expansion factor for a homogeneous spherical
perturbation is given by Eq. (62).

Direct substitution in Eqs. (60), (61) and (63) yields the ratio
of the collapse time to the current age of the (unperturbed)
universe for bound perturbations
\begin{equation}
\frac{\tau}{T_{0}}=\frac{2\int^{r_{m}/r_{1}}_{1}du\left(1-\Omega_{M}(1+\delta)-\lambda
+\Omega_{M}(1+\delta)u^{-1}+\lambda
u^2\right)^{-1/2}}{\int^{1+z_{1}}_{1}du\left(1-\Omega_{M}
-\lambda+\Omega_{M}u^{-1}+\lambda u^2\right)^{-1/2}}
\end{equation}
where $\Omega_{M}, \lambda$ are fixed at the values evaluated at
$1+z_{1}$.

\subsection{The estimation of the dynamical time}
For a particular universe, specified by $\Omega_{M0},
\lambda_{0}$, Eq. (66) defines a monotonic relation between
$\delta$ and $\tau/T_{0}$ for bound perturbations, and can
therefore be regarded as an equation for $\delta$ as a function of
$\tau/T_{0}$ (Richstone et al. 1992). Eqs. (64) and (66) specify,
for a particular $\Omega_{M0}, \lambda_{0}, \sigma$, the fraction
of the universe on some specific mass scale that has already
collapsed at cosmic time t:
\begin{equation}
F(t/T_{0})=\frac{1}{2}\left(1-erf\left(\frac{\delta(t/T_{0})}{\sqrt{2}\sigma}\right)\right)
\end{equation}
where $erf$ is the error function.

Eq. (67) predicts the fraction of the universe currently in
virialized clusters with a mass scale of rich Abell clusters. That
fraction is the product of the number density $<n>$ and mass of
rich clusters M, divided by the mean density of the universe. The
value of $\sigma$ must be found by solving Eq. (68):
\begin{equation}
\frac{1}{2}\left(1-erf\left(\frac{\delta(1)}{\sqrt{2}\sigma}\right)\right)=\frac{<n>M}{\rho_{cr}\Omega_{M0}}
\end{equation}
The simplification of Eq. (66) is useful from an analytical point
of view in order to find the function $\delta(\tau/T_{0})$. A good
approximation can be found by assuming
\begin{equation}
\tau(\delta)\propto \delta^{-3/2}
\end{equation}
The function $\tau/T_{0}$ computed from Eq. (66) is shown in Fig.
11.
 \begin{figure}[h]
 \centering
 \includegraphics[angle=0, width=8cm]{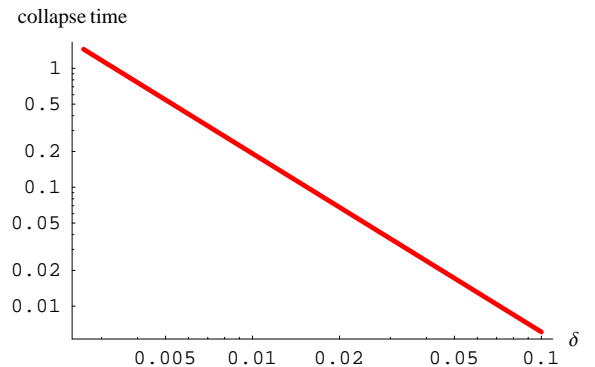}
 \caption{Collapse time $\tau/T_{0}$ as a function of the density perturbation $\delta$.}
 \end{figure}

The fraction of the mass of present day clusters that have
collapsed within the time interval $(T_{0}-\delta t, T_{0})$ given
in Fig. 12 is
\begin{equation}
\delta\tilde{F}\left(\frac{\delta
t}{T_{0}}\right)=\frac{F(1)-F(1-\delta t/T_{0})}{F(1)}
\end{equation}

\begin{figure}[h]
\centering
\includegraphics[angle=0, width=8cm]{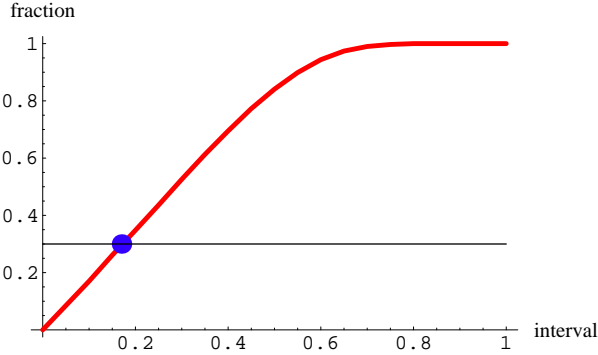}
\caption{The fraction of the mass of present day clusters that
have collapsed within the last time interval $\delta t/T_{0}$.}
\end{figure}

Forman \& Jones (1994) have studied the X-ray surface brightness
of 250 clusters of galaxies observed with Einstein. They concluded
that 30\% of that sample had X-ray substructure. Taking into
account this observational constraint one can derive the dynamical
time, and we find $\delta t/T_{0}=0.17$ (the big point in Fig.
12).

\subsection{The retarding of an X-ray substructure}
Since Eq. (51) can be solved analytically, we estimate the
retarding of a X-ray substructure from this equation when the
characteristic parameter $\lambda^2/\omega^2_{0}>>1$.\\
The solution of Eq. (51) in case $\lambda/\omega_{0}>1$ with
initial conditions ($x(0)=-R, v(0)=0$) is
\begin{eqnarray}
&&x=(C_{1}+C_{2})\exp(-(\lambda-\sqrt{\lambda^2-\omega^2})\cdot t)
+\\ &&+(C_{1}-C_{2})\exp(-(\lambda+\sqrt{\lambda^2-\omega^2})\cdot
t)
\end{eqnarray}
where
\begin{eqnarray}
&&C_{1}=-R/2\\
&&C_{2}=-\frac{\lambda
R}{2\sqrt{\lambda^2-\omega^2_{0}}}\nonumber%+\frac{v}{2\sqrt{\lambda^2-\omega^2_{0}}}\nonumber
\end{eqnarray}
The X-ray substructure velocity is found from the Eq. (71):
\begin{equation}
\frac{dx}{dt} = A e^{-\lambda t}
ch(\sqrt{\lambda^2-\omega_{0}^2}\cdot t) + B e^{-\lambda t}
sh(\sqrt{\lambda^2-\omega^2_{0}}\cdot t)
\end{equation}
where
\begin{eqnarray}
&&A=2 (-\lambda C_{1} +\sqrt{\lambda^2-\omega_{0}^2} C_{2})\\
&&B=2 (-\lambda C_{2} +\sqrt{\lambda^2-\omega_{0}^2}
C_{1})\nonumber
\end{eqnarray}
If the characteristic parameter $\lambda^2/\omega^2_{0}\gg1$, we
have
\begin{eqnarray}
&&\exp(-\lambda t + \sqrt{\lambda^2-\omega_{0}^2} t)\approx
\exp\left(-\frac{\omega_{0}^2}{2\lambda} t\right)\\
&&\exp(-\lambda t -\sqrt{\lambda^2-\omega^2} t)\approx
\exp(-2\lambda t)\nonumber
\end{eqnarray}
Therefore when the time $t\gg1/\lambda$, we can consider only the
terms of Eq. (73) that contain the factor $\exp(-\lambda t
+\sqrt{\lambda^2-\omega_{0}^2} t)$. The dependence of the velocity
with time $(t\gg1/\lambda)$ is
\begin{equation}
\frac{dx}{dt}\approx
R\frac{\omega^2_{0}}{2\lambda}\exp\left(-\frac{\omega^2_{0}}{2\lambda}
t\right)
\end{equation}
The derived velocity of the X-ray substructure in case
$\lambda/\omega_{0}>1$ is much smaller than the dark halo velocity
$(\sim R\omega_{0})$. Therefore for merger events, when this
characteristic parameter is much higher than 1, the dynamical time
is much higher than the gravitational free-fall time. This
argument can explain that the merger duration is higher than the
gravitational free-fall time for many merger events.

\section{Conclusions}

There are many proofs of evidence that X-ray substructures move
with a supersonic velocity in the ICM of clusters. In their recent
review of shocks and cold fronts in galaxy clusters, Markevitch \&
Vikhlinin (2007) nicely summarize our present knowledge on
physical mechanisms taking place when clusters merge, such as
those discussed in the present paper. The analytical study of the
X-ray substructure motion is proposed in this article. The
retarding force must be taken into account to obtain a correct
solution. One of the most obvious manifestations of its importance
is the spatial segregation of the stellar component and the X-ray
substructure in the galaxy cluster 1E0657-558. The behavior of the
X-ray substructure depends on the characteristic parameter (see
Fig. 4) in the ``piston in a tube" model. When the characteristic
parameter is smaller than 1, the X-ray substructure can pass
through the main cluster core (e.g. 1E0657-558). We also examine
the case when the characteristic parameter is higher than 1 and
the effect of the retarding force is crucial (e.g. Abell~1763).
The shock velocity in front of the X-ray substructure is
calculated.  The method for predicting the position of the dark
matter halo in clusters, where the observed baryons and the
inferred dark matter are spatially segregated, is considered.

Richstone et al. (1992) used the dynamical time of the
substructure to constrain $\Omega_{M}$, but their constraint
$\Omega_{M}>0.5$ does not correspond to the current data. A
possible reason for this is that substructure lasts significantly
longer than a crossing-time in clusters. The retarding force
influences the substructure dynamical time, which is then higher
than the crossing-time. Consequently, an accurate description of
substructure propagation is a requisite to estimate $\Omega_{M}$.

\begin{acknowledgements}
The authors thank the referee for useful comments, and G.B. Lima
Neto and M. Markevitch for discussions. They are very grateful to
D.~Fadda for providing the optical image of Abell~1763 and to
M.~Montessuit and G.~Bou\'e for their help in analyzing this
image.
\end{acknowledgements}


\begin{thebibliography}{99}
\bibitem{derm}
Berrington, R.C., Dermer, C.D. 2003, ApJ, 594, 709
\bibitem{bradac}
Bradac, M., Clowe, D., Gonzales, A.H. et al. 2006, Ap.J. 652, 937
\bibitem{buote1995}
Buote, D. A., \& Tsai, J. C. 1995, ApJ, 452, 522
\bibitem{buote1997}
Buote, D. A., \& Tsai, J. C. 1997, MNRAS, 284, 439
\bibitem{buote1998}
Buote, D. A. 1998, MNRAS, 293, 381
\bibitem{buote2002}
Buote, D. A. 2002, in Mergering Processes in Galaxy Cluster, ed.
L. Feretti, I. M. Gioia, \& G. Giovannini (Dordrecht: Kluwer), 79,
astro-ph/0106057
\bibitem{caval}
Cavaliere, A., Fusco-Femiano, R. 1976, A\&A, 49, 137
\bibitem{clowe}
Clowe, D., Bradac, M., Gonzalez, A.H. et al. 2006, ApJ Letters, 648, 109
\bibitem{cola}
Colafrancesco, S. 2006, Chin. J. Astronom. Astrophys. Vol. 6, 95
\bibitem{cola2007}
Colafrancesco, S., de Bernardis, P., Masi, S., Polenta, G., Ullio,
P. 2007, A\&A 467, L1
\bibitem{dutta}
Dutta, S. N. 1995, MNRAS, 276, 1109
\bibitem{Evrard}
Evrard, A. E., Mohr, J. J., Fabricant, D. G., \& Geller, M. J.
1993, ApJ, 419, L9
\bibitem{gabi}
Gabici, S., Blasi, P., 2003, ApJ, 583, 695
\bibitem{gerbal}
Gerbal, D., Durret, F., Lachi\`eze-Rey, M., Lima Neto, G.B 1992,
A\&A, 262, 395
\bibitem{gomez}
Gomez, P.L., Pinkney, J., Burns, J. O. et al. 1997, ApJ, 474, 580
\bibitem{gun}
Gunn, J.E., \& Gott, J.R. 1972, ApJ, 176, 1
\bibitem{forman}
Forman, W., Jones, C. 1994, in Cosmological Aspects of X-Ray
Clusters of Galaxies, ed. W.C. Seitter (Kluwer)
\bibitem{hayashi}
Hayasi, E. \& White, S. D. M. 2006, MNRAS, 351, 423
\bibitem{jones}
Jones, C., Forman, W. 1999, ApJ, 511, 65
\bibitem{kolokotronis}
Kolokotronis, V., Basilakos, S., Plionis, M., \& Georgantopoulos,
I. 2001, MNRAS, 320, 49
\bibitem{lan}
Landau, L.D., \& Lifshitz, E.M. 1959, Fluid Mechanics
(Addison-Wesley Reading)
\bibitem{lima}
Lima Neto, G.B., Durret, F. 2007, A\&A, submitted
\bibitem{max2002}
Markevitch, M., Gonzalez, A.H., David, L. et al. 2002, ApJ, 567,
L27
\bibitem{max2006}
Markevitch, M. 2006, ESA SP-604: The X-ray Universe 2005, 723
\bibitem{max2007}
Markevitch, M. \& Vikhlinin, A. 2007, Physics Reports 443,  1
\bibitem{melott}
Melott, A., Chambers, S. W., Miller, C. J. 2001, ApJ, 546, 100
\bibitem{milgrom}
Milgrom, M. 1983, ApJ, 270, 365
\bibitem{milosavljevic}
Milosavljevic, M., Koda, J., Nagai, D., Nakar, E., Shapiro, P.R.
2007, ApJ 661, L131
\bibitem{mohr}
Mohr, J. J., Evrard, A. E., Fabricant, D. G., \& Gelle, M. J.
1995, ApJ, 447, 8
\bibitem{navarro}
Navarro, J.F., Frenk, C.S., White, S.D.M. 1997, ApJ, 490, 493
\bibitem{plionis}
Plionis, M. 2002, ApJ, 572, 67
\bibitem{silk}
Pointecouteau, E., Silk, J. 2005, MNRAS, 364, 654
\bibitem{randall}
Randall, S.W., Markevitch, M., Clowe, D., Gonzales, A.H., Bradac,
M. 2007, astro-ph/07040261
\bibitem{rich}
Richstone, D., Loeb, A., Turner, E.L. 1992, ApJ, 393, 477
\bibitem{spergel}
Spergel, D.N., Bean, R., Dor\'e, O. 2007, ApJS 170, 377
\bibitem{springel}
Springel, V. \& Farrar, G. 2007, ApJ submitted, astro-ph/0703232
\bibitem{suwa}
Suwa, T., Habe, A., Yoshikawa, K., \& Okamoto, T. 2003, ApJ, 588,
7
\bibitem{valdarnini}
Valdarnini, R., Ghizzardi, S., \& Bonometto, S. 1999, New
Astronomy, 4, 71
\bibitem{zwicky}
Zwicky, F. 1933, Helv. Phys. Acta, 6, 110

\end{thebibliography}
\end{document}